\documentclass[reprint, aps, prl, superscriptaddress]{revtex4-1}

\usepackage{graphicx}
\usepackage{subfigure}
\usepackage{amsmath,bm}
\usepackage{color}
\usepackage{gensymb}
\usepackage{amsfonts}
\usepackage{times}
\usepackage{makecell}
\usepackage{booktabs}
\usepackage{multirow}
\usepackage{comment}

\DeclareMathSymbol{\shortminus}{\mathbin}{AMSa}{"39}

\begin{document}

\title{Thermodynamics of emergent structure in active matter}

\author{\firstname{Emanuele} \surname{Crosato}}\email{emanuele.crosato@sydney.edu.au}
\affiliation{Complex Systems Research Group and Centre for Complex Systems, Faculty of Engineering and IT, The University of Sydney, Sydney, NSW 2006, Australia.}
\affiliation{CSIRO Data61, PO Box 76, Epping, NSW 1710, Australia.}
\author{\firstname{Mikhail} \surname{Prokopenko}}
\affiliation{Complex Systems Research Group and Centre for Complex Systems, Faculty of Engineering and IT, The University of Sydney, Sydney, NSW 2006, Australia.}
\author{\firstname{Richard E.} \surname{Spinney}}
\affiliation{Complex Systems Research Group and Centre for Complex Systems, Faculty of Engineering and IT, The University of Sydney, Sydney, NSW 2006, Australia.}

\date{\today}

\begin{abstract}
Active matter is rapidly becoming a key paradigm of out-of-equilibrium soft matter exhibiting complex collective phenomena, yet the thermodynamics of such systems remain poorly understood.
In this letter we study the nonequilbrium thermodynamics of large scale active systems capable of motility-induced phase separation and polar alignment, using a fully under-damped model which exhibits hidden entropy productions not previously reported in the literature.
We quantify steady state entropy production at each point in the phase diagram, revealing characteristic dissipation rates associated with the distinct phases and configurational structure.
This reveals sharp discontinuities in the entropy production at phase transitions and facilitates identification of the thermodynamics of micro-features, such as defects in the emergent structure.
The interpretation of the time reversal symmetry in the dynamics of the particles is found to be crucial.
\end{abstract}

\maketitle


Active matter consists of particles that can consume stored free energy reserves in order to self-propel, and as such are characteristically out-of-equilibrium~\cite{ramaswamy2010mechanics, marchetti2013hydrodynamics, bechinger2016active, ramaswamy2017active}.
Examples encompass a wide range of systems, including self-catalytic colloidal suspensions~\cite{bialke2015active}, swimming bacteria~\cite{czirok2001theory, sokolov2009enhanced}, migrating cells~\cite{szabo2006phase} and animal groups~\cite{parrish2002self, buhl2006disorder, cavagna2010scale}.
Self-propulsion, in combination with interactions amongst the particles, can give rise to non-trivial collective dynamics not observed in matter at thermal equilibrium, such as gathering, swarming and swirling~\cite{vicsek2012collective}.

Widely used models of active particles include Active Brownian Particles (ABPs)~\cite{schimansky1995structure} and Active Ornstein-Uhlenbeck Particles (AOUPs)~\cite{fodor2016far}.
Collective motion and kinetic phase transitions can be observed in such models, with the introduction of volume exclusion, e.g., between two-dimensional discs~\cite{marchetti2016minimal}.
Indeed, systems of both ABPs and AOUPs have been shown to exhibit motility-induced phase separation (MIPS), where the particles arrange themselves into regions of high and low density~\cite{fily2012athermal, redner2013structure, redner2013reentrant, cates2015motility, solon2015pressure, marchetti2016minimal, fodor2016far, cugliandolo2017phase, digregorio2018full, siebert2018critical}.
In addition, a recent study has shown that ABPs with alignment interactions can exhibit polar collective motion (or `flocking')~\cite{martin2018collective}.

Determining the phase diagrams for such behavior has been an active area of research~\cite{fily2012athermal, redner2013structure, redner2013reentrant, kapfer2015two, cugliandolo2017phase, martin2018collective, digregorio2018full, siebert2018critical}, however, there has been less focus on the thermodynamics, especially on the nonequilibrium character of the different kinetic phases, despite some progress in related field-theoretic models \cite{nardini2017entropy}.
For molecular approaches, Fodor et al.~\cite{fodor2016far} investigated the entropy production in a system of AOUPs with no alignment interactions, arguing that in a harmonic trap the dynamics respect detailed balance such that the system is in an effective equilibrium.
Later, Mandal et al.~\cite{mandal2017entropy} demonstrated that when a different definition of entropy production is used the nonequilibrium character can be recovered.
Recently, Shankar et al.~\cite{shankar2018hidden} investigated the `hidden' components of the entropy production observed only in under-damped descriptions of the particles' translational dynamics, reporting a key dependence on the time reversal symmetry (TRS) interpretation of the self-propulsion force, mirroring the distinct approaches of Fodor and Mandal.
However, the study only considered free, non-interacting particles.

In contrast, we consider a large system of ABPs interacting via volume exclusion as well as alignment enabling investigation of the entropy production associated with the emergent collective motion.
We derive expressions for the entropy production for both over and under-damped models, under odd and even interpretations of the parity of the particles' heading.
This reveals an additional hidden component due to coarse grained rotational dynamics even with under-damped translational dynamics.
Simulation of the under-damped model allows us not only to construct the phase diagram, but also quantify the steady state thermodynamics at each point in the space.
Further, we are able to examine the spatial distributions of entropy production associated with the distinct phases alongside micro-features, such as defects, in the emergent structures.

\begin{table*}[t]
\centering
\begin{tabular}{c c c} 
\Xhline{2\arrayrulewidth}
\addlinespace[0.3em]
$\displaystyle \langle\Delta \dot{S}^{\text{tot}}\rangle$ & Over-damped & Under-damped\\
\addlinespace[0.3em]
\Xhline{2\arrayrulewidth}
\addlinespace[0.3em]
TRS-odd heading
&
\makecell{
$\displaystyle
\textstyle\sum_{a=1}^N
\Big(
\beta(I\gamma_\text{R})^{\shortminus 1}\big\langle{\cal T}_a^2(\mathbf{r}, \theta)\big\rangle
+ (I\gamma_\text{R})^{\shortminus 1}\big\langle\partial_{\theta_a}{\cal T}_a(\mathbf{r}, \theta)\big\rangle
$\\
$\displaystyle
+ \textstyle\sum_{j=1}^2
\beta v_0\big\langle\partial_{r_a^j}U_a(\mathbf{r})\hat{e}^j(\theta_a)\big\rangle
\Big)
$
}
&
\makecell{
$\displaystyle
\textstyle\sum_{a=1}^N
\Big(
\gamma_\text{R}\big(\beta I\big\langle\omega_a^2\big\rangle - 1\big)
$\\
$\displaystyle
+ \textstyle\sum_{j=1}^2
\gamma\big(\beta mv_0^2 /2 - \beta m \big\langle(v^j_a)^2\big\rangle + 1\big)
\Big)
$
}
\\
\addlinespace[0.5em]
TRS-even heading
&
\makecell{
$\displaystyle
\textstyle\sum_{a=1}^N
\Big(
\beta(I\gamma_\text{R})^{\shortminus 1}\big\langle{\cal T}^2_a(\mathbf{r}, \theta)\big\rangle
+ (I\gamma_\text{R})^{\shortminus 1}\big\langle\partial_{\theta_a}{\cal T}_a(\mathbf{r}, \theta)\big\rangle
$\\
$\displaystyle
+ \ m\gamma\beta v_0^2
- \textstyle\sum_{j=1}^2
\beta v_0\big\langle\partial_{r_a^j}U_a(\mathbf{r})\hat{e}^j(\theta_a)\big\rangle
\Big)
$
}
&
\makecell{
$\displaystyle
\textstyle\sum_{a=1}^N
\Big(
\gamma_\text{R}\big(\beta I\big\langle\omega_a^2\big\rangle - 1\big)
$\\
$\displaystyle
+ \textstyle\sum_{j=1}^2
\gamma\big(\beta m \big\langle(v^j_a)^2\big\rangle - 1 \big)
\Big)
$
}\\
\addlinespace[0.3em]
\Xhline{2\arrayrulewidth}
\end{tabular}
\caption{Expected entropy production rate for a system of interacting ABPs at steady state, described by over-damped and under-damped dynamics, and for odd and even interpretation of the particles' heading under TRS.}
\label{tab:ep}
\end{table*}


We consider a system of $N$ two-dimensional, disc-shaped ABPs of radius $R$, mass $m$,  moment of inertia $I$, self-propulsion speed $v_0$ and translational and rotational mobility coefficients $\gamma$ and $\gamma_{\text{R}}$.
The position and heading of each particle $a$ are denoted as $\mathbf{r}_a{=}\{r_a^1 , r_a^2\}$ and $\theta_ a$ respectively (variables without subscripts or superscripts are to be understood as the total set of such variables in the system, e.g. $\mathbf{r}{=}\{\mathbf{r}_1,\ldots,\mathbf{r}_N\}$).
The self-propulsion force is modeled as ${\cal P}(\theta_a){=}\{{\cal P}^1(\theta_a),{\cal P}^2(\theta_a)\} {=} m\gamma v_0\hat{e}(\theta_a)$, where $\hat{e}(\theta_a){=}$$\{\hat{e}^1(\theta_a), \hat{e}^2(\theta_a)\}$${=}\{\cos(\theta_a), \sin(\theta_a)\}$.
Excluded volume interactions are modeled using a truncated and shifted Lennard-Jones potential $U(\mathbf{r}){=}\sum_a U_a(\mathbf{r})$ with $U_a(\mathbf{r}){=}\sum_{b\ne a}\epsilon[(2R/\mathbf{r}_{ab})^{12} {-} (4R/\mathbf{r}_{ab})^6] + \epsilon$ if $|\mathbf{r}_{ab}|{\le}R$ and $U_a(\mathbf{r}) {=} 0$ if $|\mathbf{r}_{ab}|{>}R$, where $\epsilon$ is the depth of the potential well and $\mathbf{r}_{ab}{=}\mathbf{r}_a{-}\mathbf{r}_b$.
Finally, informed by the Kuramoto model~\cite{acebron2005kuramoto}, alignment interactions are modeled as ${\cal T}_a(\mathbf{r}, \theta) {=} {-}K\sum_{b\ne a} g(\mathbf{r}_{ab})\sin(\theta_a {-} \theta_b)$ where $K$ is the coupling strength and $g(\mathbf{r}_{ab}){=}1$ if $|\mathbf{r}_{ab}|{\le}2R$ and zero otherwise.

A minimal description of the system is given by the following (over-damped) stochastic differential equations (SDEs):
\begin{align}
\begin{split}
dr_a^j =& \ (m \gamma)^{\shortminus 1}{\cal F}^j_a(\mathbf{r}, \theta) dt + \sqrt{2/\beta m\gamma}dW_{r_a^j} ,
\end{split}\\
\begin{split}
d\theta_a =& \ (I \gamma_{\text{R}})^{\shortminus 1}{\cal T}_a(\mathbf{r}, \theta) dt + \sqrt{2/\beta I\gamma_{\text{R}}}dW_{\theta_a} ,
\end{split}
\end{align}
where ${\cal F}^j_a(\mathbf{r}, \theta) {=} {\cal P}^j(\theta_a) {-}\partial_{r^j_a}U_a(\mathbf{r})$, $i$ is the spatial dimension, $\beta$ is the inverse temperature (with units $k_B{=}1$) and $W_{r^j_a}$ and $W_{\theta_a}$ are uncorrelated Wiener processes, such that $\langle dW_{r^j_a}dW_{\theta_a}\rangle{=}0$, $\langle dW_{\theta_a}dW_{\theta_b}\rangle{=}\delta_{ab}dt$ and $\langle dW_{r^j_a}dW_{r^k_b}\rangle{=}\delta_{jk}\delta_{ab}dt$.
A finer grained description of the dynamics includes the translational and rotational velocities $\mathbf{v}_a{=}\{v_a^1 ,  v_a^2\}$ and $\omega_a$ through the under-damped SDEs:
\begin{align}
\begin{split}
dr_a^j =& \ v_a^j dt ,
\end{split}
\label{eq:under1}\\
\begin{split}
dv_a^j =&\ {-}\gamma v_a^j dt + m^{\shortminus 1}{\cal F}^j_a(\mathbf{r}, \theta) dt + \sqrt{2\gamma/\beta m}dW_{v_a^j} ,
\end{split}\\
\begin{split}
d\theta_a =& \ \omega_a dt ,
\end{split}\\
\begin{split}
d\omega_a =&\ {-}\gamma_{\text{R}}\omega_a dt + I^{\shortminus 1}{\cal T}_a(\mathbf{r}, \theta) dt + \sqrt{2\gamma_{\text{R}}/\beta I}dW_{\omega_a} ,
\end{split}
\label{eq:under2}
\end{align}
where $W_{v_a^j}$ and $W_{\omega_a}$ are also independent Wiener processes.


To understand the thermodynamics of these models we may turn to the framework of stochastic thermodynamics \cite{seifert2008stochastic}.
A central quantity of interest is the steady dissipation, or entropy production, which in such a formalism can be interpreted as a measure of dynamical irreversibility.
Taking $k_B{=}1$ and defining $\Omega {=} \{\mathbf{r}, \mathbf{v}, \theta, \omega\}$ (or $\Omega {=} \{\mathbf{r}, \theta\}$ for the over-damped system) as the total state of the system, the entropy production of an individual realization $\vec{\Omega}{=}\{\Omega(t)|t{\in}[t_0,\tau]\}$, over the interval $[t_0,\tau]$, is given by $\Delta S^\text{tot} {=} \ln(P[\vec{\Omega}]/P^\dagger[\vec{\Omega}^\dagger])$~\cite{seifert2005entropy, seifert2008stochastic}. Here ${P}$ and ${P}^\dagger$ are the probability measures for the forward and time reversed dynamics, $\vec{\Omega}^\dagger{=}\{\Omega^\dagger(t)|t{\in}[t_0,\tau]\}$ and $\Omega^\dagger(t){=}\mathbf{\varepsilon}\Omega(\tau {+}t_0{-}t)$ where $\mathbf{\varepsilon}$ is a time reversal operator~\cite{spinney2012nonequilibrium}.
Consequently, the entropy production is equal to the log ratio of the likelihood of a given trajectory against its time reverse.
For stationary, autonomous and time symmetric dynamics, e.g. ABPs in a steady state, ${P}^\dagger{=}{P}$.

The total entropy production, comprising the change in entropy of the system and the environment, obeys an integral fluctuation theorem $\langle\exp[-\Delta S^\text{tot}]\rangle{=}1$.
Thus the strict inequality $\langle\Delta S^\text{tot}\rangle {\geq} 0$ holds by Jensen's inequality, characterizing the second law.
For SDEs, expressions for the total entropy production can be found exactly given knowledge of the probability density functions over the variables~\cite{spinney2012entropy}, whilst expressions for the environmental entropy production can be determined in terms of the trajectories only.
In the steady state, however, the expected change of system entropy vanishes and the mean medium entropy production is equal to the mean total entropy production allowing empirical calculation of expectations without the need for solving the associated Fokker-Planck equation.

Utilizing the formalism in~\cite{spinney2012entropy} we derive the expected medium entropy production for the total system of ABPs, assuming a steady state, for both over-damped and under-damped dynamics.
The results depend crucially on the operator $\mathbf{\varepsilon}$, with uncertainty in the literature as to the time reversal symmetry of the particles' orientation~\cite{shankar2018hidden}.
These entropy productions, under both odd and even interpretations of $\theta$ are reported in Table~\ref{tab:ep} with details in the Supplemental Material (SM).
These expressions are quite general, however, in the absence of alignment, external, and exclusion interactions, such that ${\cal T}_a(\mathbf{r}, \theta){=}0$ and $\partial_{r_a^j}U(\mathbf{r}){=}0$, we can recover and generalize the results for free ABPs in~\cite{shankar2018hidden}.
The over-damped results follow directly, however, results using under-damped translational dynamics depend on the treatment of the rotational degrees of freedom.
Generally the individual free particle entropy production takes the form
\begin{equation}
\langle\Delta\dot{S}^\text{tot}\rangle =m\beta\gamma v_0^2(1-\gamma^2\mathcal{G})
\end{equation}
for the odd interpretation of $\theta$ and
\begin{equation}
\langle\Delta\dot{S}^\text{tot}\rangle =m\beta\gamma^3v_0^2\mathcal{G}
\end{equation}
for the even interpretations of $\theta$, where
\begin{equation}
\mathcal{G}=\lim_{t\to\infty}2\int_0^tdt_1\int_0^tdt_2 \;e^{\shortminus\gamma(2t\shortminus t_1\shortminus t_2)}\langle \hat{e}^1(t_1)\hat{e}^1(t_2)\rangle.
\end{equation}

Shankar et al.~\cite{shankar2018hidden}, whilst considering under-damped translational motion, utilize over-damped rotational motion with $2\langle\hat{e}^1(t_1)\hat{e}^1(t_2)\rangle{=}e^{(\beta I \gamma_\text{R})^{-1}|t_1\shortminus t_2|}$ and thus $\mathcal{G}_\text{over}{=}(\gamma(\gamma{+}(\beta I\gamma_\text{R})^{\shortminus 1}))^{\shortminus1}$.
However, a fully under-damped description of the free-particle dynamics yields $2\langle\hat{e}^1(t_1)\hat{e}^1(t_2)\rangle{=}e^{(\beta I \gamma^2_R)^{\shortminus 1}(1\shortminus\gamma_\text{R}|t_1\shortminus t_2|\shortminus\exp[\shortminus\gamma_\text{R}|t_1\shortminus t_2|])}$.
Except for specific choice of parameters (e.g., all free parameters set to $1$), the integral has no closed form solution, but strictly satisfies $\mathcal{G}_\text{under}\!\geq\!\mathcal{G}_\text{over}$ indicating an additional hidden component in the entropy production (see details in the SM).

This illustrates the well known property that contributions are lost through coarse graining procedures~\cite{esposito2012stochastic}.
Such absent terms have been referred to as `anomalous'~\cite{celani2012anomalous} or `hidden' and have been previously implicated in heat transfer where under-damped models are crucial in order to observe physically plausible entropy productions \cite{spinney2012entropy}. 
However, here the results are particularly nuanced as \emph{translational} entropy productions in the free particle results are hidden due to coarse-graining in the \emph{rotational} degrees of freedom and the discrepancies are non-trivial.
For instance, setting all free parameters to $1$ yields $\mathcal{G}_\text{under}{=}(2e-4)\mathcal{G}_\text{over}\!\simeq\!1.44\;\mathcal{G}_\text{over}$, with commensurate over and under estimates in the entropy production for odd and even $\theta$ respectively.
In light of this, despite apparent additional complication, we proceed utilizing the fully under-damped model so that we can be assured no features are either missing or are introduced as artifacts.

Importantly, as the Wiener processes are assumed to be uncorrelated (also known as a bipartite, or rather multipartite property ~\cite{spinney2018entropy}), we may associate entropy productions with \emph{individual} particles, with the total being their sum.
For the odd interpretation the expected entropy production rate in the medium for particle $a$ (no longer assuming a steady state) is
\begin{equation}
\begin{aligned}
\langle\Delta \dot{S}^\text{med}_a|\Omega\rangle &=
\gamma_\text{R}\left(\beta I\left\langle\omega_a^2\right\rangle {-} 1\right)\\
&\quad+{\textstyle \sum_{j=1}^2 }{\gamma}\left(\beta m \langle(v^j_a {-}v_0\hat{e}^j(\theta_a))^2\rangle {-} 1 \right),
\label{eq:ep-noeq-odd}
\end{aligned}
\end{equation}
where again $k_B{=}1$.
For the even interpretation we have
\begin{align}
\langle\Delta \dot{S}^\text{med}_a|\Omega\rangle &\!=\!
\gamma_\text{R}\!\left(\beta{I}\!\left\langle\omega_a^2\right\rangle\!{-} 1\right)\!{+}{\textstyle\sum_{j=1} ^{2}}{\gamma} \big(\beta{m}\langle\!{v^j_a}^2\rangle {-} 1 \big).
\label{eq:ep-noeq-even}
\end{align}
The distinction between the parity interpretations is striking: under the even parity interpretation, the entropy production is manifestly a measure of the deviation away from equipartition expected at thermodynamic equilibrium in both the translational and rotational degrees of freedom.
In contrast, under the odd parity interpretation the entropy production arising from the translational variables is modified such that it quantifies deviation from an effective equipartition, relative to the instantaneous heading and typical speed.


\begin{figure}[t]
\centering
\includegraphics[trim={58mm 30mm 25mm 28mm}, clip, width=1\columnwidth]{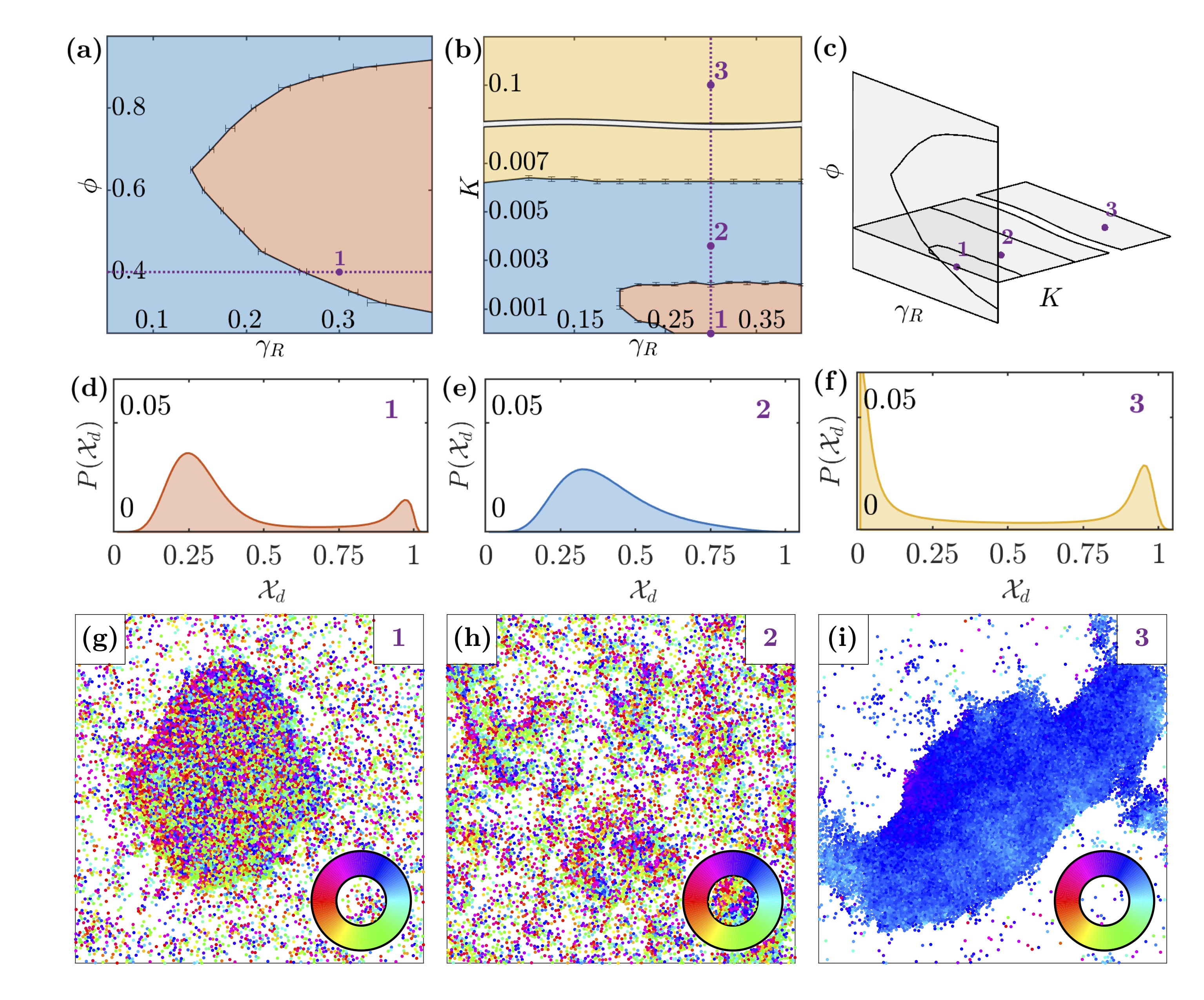}
\caption{
Summary of the kinetic phases.
(a) shows the phase diagram of the system with respect to $\phi$ and $\gamma_{\text{R}}$, when $K{=}0$.
(b) shows the phase diagram with respect to $\gamma_{\text{R}}$ and $K$, at $\phi{=}0.4$.
(c) illustrates the two sections through the $\phi{\shortminus}K{\shortminus}\gamma_{\text{R}}$ space.
In both diagrams the error bars indicate the intervals within which the phase transitions are observed to occur, based on the simulations.
The black lines are approximations of the critical lines, given the error bars.
The purple lines represent trajectories across the phase diagrams over which the expected steady state entropy production rate is shown in Fig.\ref{fig:ep}.
Three representative points along these lines, corresponding to the three phases, are labelled with numbers.
For each of them, (d-f) show the distribution of the local density ${\cal X}_d$ (with $d{=}4.5$), while (g-i) show a typical configuration observed during the simulations (color represents the particles' heading).
}
\label{fig:phase-diagrams}
\end{figure}

\begin{figure}[t]
\centering
\includegraphics[width=1\columnwidth]{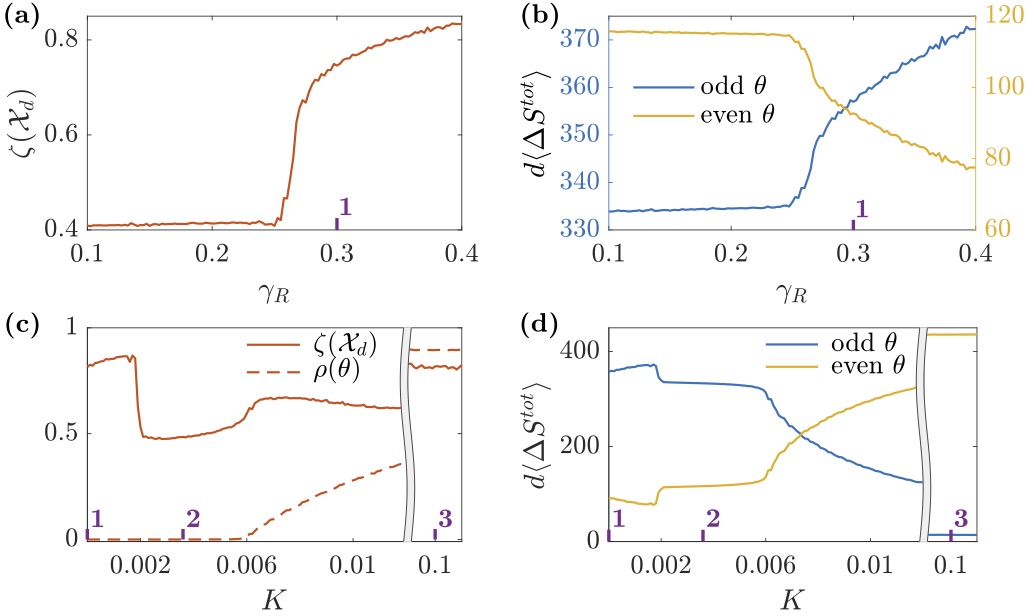}
\caption{
Expected steady state entropy production rate over the tree kinetic phases.
In (a) and (b) $\phi{=}0.4$ and $K{=}0$, while $\gamma_{\text{R}}$ is varied (cf. purple line in Fig~\ref{fig:phase-diagrams}(a)).
(a) shows the average bimodality coefficient $\zeta({\cal X}_d)$ (with $d{=}4.5$) at steady state, while (b) shows the expected entropy production rate for both the odd and even interpretation of $\theta$.
In (c) and (d) $\phi{=}0.4$ and $\gamma_{\text{R}}{=}0.3$, while $K$ is varied (cf. purple line in Fig~\ref{fig:phase-diagrams}(b)).
(c) shows the average $\zeta({\cal X}_d)$ and the average alignment coefficient $\rho(\theta_a)$ at steady state, while (d) shows the expected entropy production rate for the odd and even interpretation of $\theta$.
In all figures, the purple ticks indicate the representative points (cf. Fig.~\ref{fig:phase-diagrams}).
}
\label{fig:ep}
\end{figure}

The system is simulated by integrating Eqs.~(\ref{eq:under1}-\ref{eq:under2}) with a stochastic velocity Verlet algorithm (details can be found in the SM) and we explore its behavior over $\gamma_{\text{R}}$, $K$ and the particles density $\phi$.
These variables were chosen specifically to investigate the thermodynamic character of the emergent structures, rather than those which derive from the strength of the self-propulsion force and external heat bath, which would together entirely determine the entropy production of a free particle without the rotational degree of freedom.
For instance, MIPS is typically controlled using the P\'{e}clet number $\text{Pe}\!\propto\! v_0\beta\sqrt{mI\gamma\gamma_{\text{R}}}$~\cite{bechinger2016active} by varying the propulsion force, environmental temperature and relative timescales.
Instead, we restrict ourselves to varying only the relative timescales through $\gamma_\text{R}$.
Consequently, we hold all other variables constant, setting $N{=}10000$, $R{=}0.5$, $v_0{=}3$, $m{=}I{=}\gamma{=}1$, $\beta{=}50$ and $\epsilon{=}1$, and also utilize periodic boundary conditions.

In order to characterize the configurational change associated with MIPS we utilize the local (per particle) sixfold bond-orientational order: $\big|q_6(a)\big| {=} \left|\frac{1}{6} \sum_{b\in{\cal N}_a} e^{i6\alpha_{ab}}\right|$, where $\alpha_{ab}$ is the angle between $\mathbf{r}_{ab}$ and an arbitrary axis and ${\cal N}_a$ are the closest $6$ neighboring particles of $a$.
An order parameter for the phase separation is therefore provided by the average bond-orientational order $\langle|q_6(a)|\rangle$.
This can be complemented by statistics of the local density ${\cal X}_d$, defined as the empirical density within a radius $d$, since we expect a bimodal distribution under MIPS.
We consider the bimodality coefficient $\zeta({\cal X}_d) {=} (\lambda({\cal X}_d) + 1)/\kappa({\cal X}_d)$, where $\lambda({\cal X}_d)$ and $\kappa({\cal X}_d)$ are, respectively, the third and and the fourth standardized moments of ${\cal X}_d$.
The alignment within the system is instead quantified as $\rho(\theta) {=} \langle 2 \cos^2(\theta_a {-} \bar{\theta}) {-} 1 \rangle$, where $\bar{\theta}$ is the mean heading across all particles.
We also introduce a measure of per particle alignment $\tilde{\rho}_a(\mathbf{r}, \theta) {=} \langle 2 \cos^2(\theta_a {-} \bar{\theta}_{{\cal N}_a}) {-} 1 \rangle$, where $\bar{\theta}_{{\cal N}_a}$ is the mean heading within ${\cal N}_a$.

When only excluded volume interactions are considered (i.e., $K{=}0$) as expected we observe two distinct phases: a phase with MIPS and a phase without MIPS, separated by a single critical value of $\gamma_{\text{R}}$ for any given $\phi$ (see Fig.~\ref{fig:phase-diagrams}(a)).
Analogous behavior was observed in~\cite{digregorio2018full, siebert2018critical}.
A third kinetic phase is possible when alignment interactions are included, characterized by both polar order and MIPS (see Fig.~\ref{fig:phase-diagrams}(b)).
At density $\phi{=}0.4$, for example, this third phase is observed for values of $K\!\gtrsim\!0.006$.
For lower values of $K$ the system does not exhibit polar order, however the alignment interactions affect MIPS, which occurs only at values of $K\!\lesssim\!0.002$.
Importantly, the two MIPS phases with and without polar order are emergent via two distinct and incompatible mechanisms, both having distinct effects upon the thermodynamics.
Explicitly, phase separation without polar order arises due to long rotational correlation times which induces jamming-like behavior, whilst phase separation with polar order arises due to flocking behavior.
At intermediate $K$ there is enough alignment to reduce the correlation times of the single particle rotational dynamics, but not enough to cause global rotational correlations necessary for flocking.


The steady state, nonequilibrium, thermodynamics of the three kinetic phases is illustrated by considering two representative trajectories through the phase diagram.
The first follows the onset of MIPS in the absence of alignment interactions (i.e., $K{=}0$) at fixed density $\phi{=}0.4$ by varying $\gamma_\text{R}$ indicated in Fig.~\ref{fig:phase-diagrams}(a).
The structural and thermodynamic character along the trajectory is then illustrated in Fig.~\ref{fig:ep}(a-b): increasing $\gamma_{\text{R}}$ up to the critical value $\sim\!0.26$ has little effect before an abrupt increase in the bimodality coefficient at the critical point indicating the onset of MIPS.
This is accompanied by a decrease in mean particle velocity through jamming causing an equally abrupt change in the expected steady state entropy production.
Crucially, odd and even TRS imply completely opposite variation in the entropy production rates with an even interpretation implying lower dissipation under MIPS and vice versa.
This is a qualitative distinction in the thermodynamics associated with collective motion and emergent structure, not manifest in the free particle dissipation~\cite{shankar2018hidden}.

The second trajectory is indicated in Fig.~\ref{fig:phase-diagrams}(b) for $\phi{=}0.4$ and $\gamma_{\text{R}}{=}0.3$ as MIPS without polar order is first interrupted and then reintroduced with polar order by increasing the alignment interactions through $K$.
The relevant structural and thermodynamic consequences are then illustrated in Fig.~\ref{fig:ep}(c-d).
Polar order, measured through $\rho$, emerges beyond a critical $K{\sim}0.006$.
However, spatial order is more complicated with a large and increasing bimodality coefficient abruptly dropping when the jamming mechanism is interrupted, before distinctly rising at the onset of polar order due to flocking.
The bimodality coefficient then slowly increases, although not monotonically, as MIPS with polar order dominates.
Below the onset of polar order the entropy production follows the spatial order as in the $K{=}0$ trajectory with mean velocity controlled by jamming.
However, beyond this point the entropy production follows the polar order as the increased alignment allows for higher velocities.
Once again, odd and even TRS interpretations implicate opposite variation in the nonequilibrium behavior, with the highly aligned state corresponding to high entropy production under an even interpretation.


\begin{figure}[t]
\centering
\includegraphics[width=1\columnwidth]{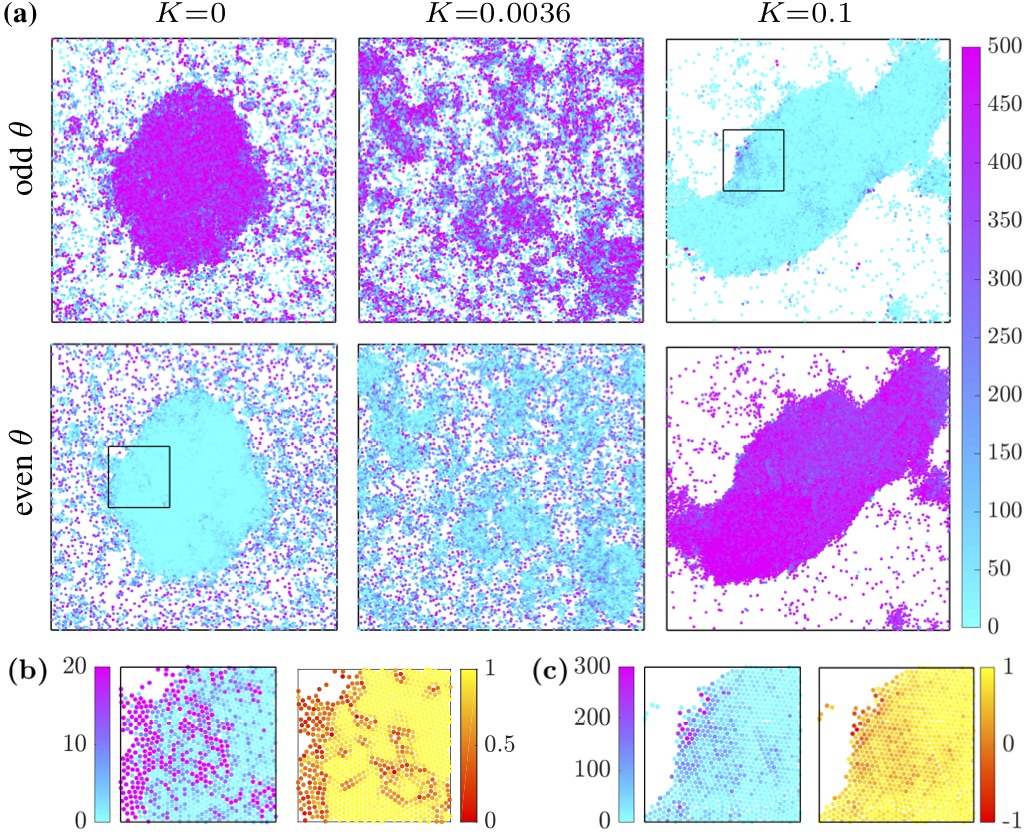}
\caption{
Expected entropy production rate associated to individual particles.
(a) shows the same three configurations in Fig.~\ref{fig:phase-diagrams}(d-f) with color representing each particles' expected entropy production rate (see Eq.~\eqref{eq:ep-noeq-odd} and Eq.~\eqref{eq:ep-noeq-even}), distinguishing between odd and even interpretations of $\theta$ under TRS.
(b)-left magnifies the box in (a) corresponding to $K{=}0$ and even $\theta$, using a higher resolution for the entropy production that can capture small differences between low values.
(b)-right shows the local sixfold bond-orientational order $|q_6|$, highlighting spatial defects across the emergent structure.
(c)-left magnifies the box in (a) corresponding to $K{=}0.1$ and odd $\theta$, again using a different resolution for the entropy production.
(c)-right shows the local alignment $\tilde{\rho}$, highlighting orientational defects.
}
\label{fig:ep-screen}
\end{figure}

The spatial distribution of the entropy production can be investigated by considering the dissipation associated with individual particles (see Eq.~\eqref{eq:ep-noeq-odd} and~\eqref{eq:ep-noeq-even}).
This is exemplified in Fig.~\ref{fig:ep-screen}, for the three configurations of the system previously seen in Fig.~\ref{fig:phase-diagrams}(g-i).
In the absence of polar order the dissipative contribution from each particle closely follows the local density (Fig.~\ref{fig:ep-screen}(a), $K{=}0$ and $K{=}0.0036$).
When polar order is high (Fig.~\ref{fig:ep-screen}(a), $K{=}0.1$), this trend is reversed reflecting the distinct phase separation mechanism.
An odd interpretation suggests that non-polarized clusters are highly dissipative and polarized clusters are closer to equilibrium and vice versa under and even interpretation.

This ability to quantify the thermodynamic effects of specific local spatial configurations allows consideration of defects in the emergent structures.
In this manner we find a nonequilibrium analogue to the increased entropies of crystalline structures due to defects.
Specifically, defects are responsible for either increases or \emph{decreases} in the entropy \emph{production} depending on the phase and TRS interpretation.
These deviations can be directly associated with individual particles.
For example, Fig.~\ref{fig:ep-screen}(b) contrasts the expected entropy production rates of individual particles with their local sixfold bond-orientational order $|q_6|$ under MIPS without polar order.
In this phase, particles along the spatial defects are characterized by higher (lower) entropy production rates compared to the particles in highly ordered regions for even (odd) $\theta$.
Similarly, Fig.~\ref{fig:ep-screen}(c) contrasts the expected entropy production rates with the local alignment $\tilde{\rho}$ under MIPS with polar order.
In this phase, for suitably high $K$, polar defects (as measured by $\tilde{\rho}$) are characterized by lower (higher) entropy production for even (odd) $\theta$.


In this letter we have explored the thermodynamic character that emerges from the rich collective dynamics exhibited by active matter and highlighted a hidden entropy production where rotational timescales impact dissipation in the translational degrees of freedom.
Our results suggest that the richness, commonly associated with the phase structure of active matter, is mirrored in its thermodynamics, opening up a new tool to study collective phenomena on both a micro and macroscopic scale.
Important questions remain, including the delicate issue of TRS which we have shown to dramatically influence any thermodynamic interpretation.
We hope that the work will contribute to a deeper understanding of the thermodynamics of active systems and, more broadly, the dynamics that can lead to emergent structures.


\begin{acknowledgments}
E.C. was supported by the University of Sydney's ``Postgraduate Scholarship in the field of Complex Systems'' from Faculty of Engineering \& IT and by a CSIRO top-up scholarship.
The authors acknowledge the University of Sydney HPC service at The University of Sydney for providing HPC resources that have contributed to the research results reported within this paper.
\end{acknowledgments}


\bibliographystyle{apsrev4-1}

%


\clearpage
\newpage
\setcounter{page}{1}
\setcounter{equation}{0}

\begin{widetext}

\thispagestyle{empty}

\begin{center}
{\large\textbf{Thermodynamics of emergent structure in active matter\\\vspace{6mm}Supplemental material\\\vspace{6mm}}}

Emanuele Crosato,\textsuperscript{1,2,*} Mikhail Prokopenko,\textsuperscript{1} and Richard E. Spinney\textsuperscript{1}\\
\vspace{1mm}
\textit{
\textsuperscript{1}Complex Systems Research Group and Centre for Complex Systems,\\
Faculty of Engineering and IT, The University of Sydney, Sydney, NSW 2006, Australia.\\
\textsuperscript{2}CSIRO Data61, PO Box 76, Epping, NSW 1710, Australia.\\
(Dated: \today)}
\end{center}
\vspace{2mm}

\section*{Derivation of the entropy productions}

The over-damped and under-damped models presented in the main text involve continuous Markovian dynamics described by uncorrelated stochastic differential equations (SDEs).
Given the total phase space $\Omega$, the evolution of a single degree of freedom, ${x_a\in\{{r}_a^1, {r}_a^2, {v}_a^1, {v}_a^2, \theta_a, \omega_a\}}$, $a\in\{1,\ldots,N\}$, can be expressed in terms of a deterministic component described by the function $A_x(\Omega(t),t)$ and a stochastic component described by the function $B_x(\Omega(t),t)$ in conjunction with a Wiener process:
\begin{equation}
dx _a= A_{x_a}(\Omega(t),t)dt + B_{x_a}(\Omega(t),t)dW_{x_a}.
\end{equation}
Note, here $W_{x_a}$ are Wiener processes satisfying $dW_{i}dW_{j}=\delta_{ij}dt$.
The deterministic dynamics $A_{x_a}(\Omega(t),t)$ can be further divided into reversible and irreversible components~\cite{risken1996fokker}:
\begin{equation}
dx_a = A^\text{REV}_{x_a}(\Omega(t),t)dt + A^\text{IR}_{x_a}(\Omega(t),t)dt + B_{x_a}(\Omega(t),t)dW_{x_a} .
\end{equation}
Diffusion coefficients can then be associated to each coordinate such that $D_{x_a}(\Omega(t),t)=B_{x_a}(\Omega(t),t)^2/2$.

Following~\cite{spinney2012entropy}, noting that there is no multiplicative noise in the models, and that the dynamics in question are autonomous, we have:
\begin{equation}
\label{eq:ep}
d\Delta S^\text{med} = \sum_{a=1}^N\sum_{{x_a\in\{{r}_a^1, {r}_a^2, {v}_a^x, {v}_a^2, \theta_a, \omega_a\}}}\frac{A^\text{IR}_{x_a}(\Omega(t))}{D_{x_a}}\circ dx_a - \frac{A^\text{IR}_{x_a}(\Omega(t))A^\text{REV}_{x_a}(\Omega(t))}{D_{x_a}}dt
\end{equation}
where the $\circ$ notation indicates a Stratonovich integration rule. We also note the convention $0/0 = 0$ for deterministic co-ordinates.

\subsection*{Entropy production in the over-damped model}

First we consider the over-damped model described in the main text:
\begin{align}
\begin{split}
d{r}_a^j =& \ v_0\hat{e}^j(\theta_a) dt - \frac{1}{m\gamma}\partial_{{r}^j_a}U_a(\mathbf{r}) dt + \sqrt{\frac{2}{\beta m\gamma}}dW_{{r}_a^j} ,
\end{split}\\
\begin{split}
d\theta_a =& \ \frac{1}{I\gamma_{\text{R}}}{\cal T}_a(\mathbf{r}, \theta) dt + \sqrt{\frac{2}{\beta I\gamma_{\text{R}}}}dW_{\theta_a} .
\end{split}
\end{align}
Regardless of the paritity of $\theta$ under TRS, we have $A^{\text{IR}}_{\theta_a} {=} {\cal T}_a( \mathbf{r},\theta)/I$, $A^{\text{REV}}_{\theta_a} {=} 0$, $D_{\theta_a} {=} (\beta I\gamma_\text{R})^{-1}$ and $D_{{r}_a^j} = (\beta m\gamma)^{-1}$.


\subsubsection*{Odd self-propulsion}

For the odd interpretation of $\theta$ under TRS, we have $A^{\text{IR}}_{{r}_a^j} {=} - \partial_{{r}^j_a}U_a(\mathbf{r})/m\gamma$ and $A^{\text{REV}}_{{r}_a^j} {=} v_0\hat{e}^j(\theta_a)$.
Applying Eq.~\eqref{eq:ep} we obtain:
\begin{equation}
d\Delta S^\text{med} = 
\sum_{a=1}^N
\Bigg(
\beta{\cal T}_a( \mathbf{r},\theta) \circ d\theta_a
- \sum_{j=1}^2
\beta\bigg(
\partial_{{r}_a^j}U_a(\mathbf{r})\circ d{r}_a^j
+ v_0\hat{e}^j(\theta_a)\partial_{{r}_a^j}U_a(\mathbf{r}) dt
\bigg)
\Bigg) .
\end{equation}
After conversion to It\={o} form, in the steady-state we may assume $\langle dU(\mathbf{r})\rangle{=}0$ and $\langle(\hat{e}^j(\theta_a))^2\rangle {=} 1/2$ and thus obtain:
\begin{equation}
\frac{d\langle\Delta S^\text{med}\rangle}{dt} = 
\sum_{a=1}^N
\Bigg(
\frac{\beta\langle{\cal T}_a^2( \mathbf{r},\theta)\rangle}{I\gamma_\text{R}}
+ \frac{\langle\partial_{\theta_a}{\cal T}_a( \mathbf{r},\theta)\rangle}{I\gamma_\text{R}}
+ \sum_{j=1}^2
\beta v_0\left\langle\partial_{{r}_a^j}U_a(\mathbf{r})\hat{e}^j(\theta_a)\right\rangle
\Bigg) .
\end{equation}


\subsubsection*{Even self-propulsion}

For the even interpretation of $\theta$, we obtain $A^{\text{IR}}_{{r}_a^j} {=}  v_0\hat{e}^j(\theta_a) - \partial_{{r}^j_a}U_a(\mathbf{r})/m\gamma$ and $A^{\text{REV}}_{{r}_a^j} {=} 0$.
Applying Eq.~\eqref{eq:ep}, we then obtain:
\begin{equation}
d\Delta S^\text{med} = 
\sum_{a=1}^N
\Bigg(
\beta{\cal T}_a( \mathbf{r},\theta)\circ d\theta_a
+ \sum_{a=1}^N
\beta\Big(v_0\hat{e}^j(\theta_a) - \partial_{{r}_a^j}U_a(\mathbf{r})\Big)dt
\Bigg) .
\end{equation}
Once again converting to It\={o} form and assuming a steady-state such that $\langle dU(\mathbf{r})\rangle{=}0$ and $\langle(\hat{e}^j(\theta_a))^2\rangle {=} 1/2$, we obtain:
\begin{equation}
\frac{d\langle\Delta S^\text{med}\rangle}{dt} = 
\sum_{a=1}^N
\Bigg(
\frac{\beta\langle{\cal T}_a^2( \mathbf{r},\theta)\rangle}{I\gamma_\text{R}}
+ \frac{\langle\partial_{\theta_a}{\cal T}_a( \mathbf{r},\theta)_a\rangle}{I\gamma_\text{R}}
+ \beta m\gamma v_0^2
- \sum_{a=1}^N
\beta v_0\left\langle\partial_{{r}_a^j}U_a(\mathbf{r})\hat{e}^j(\theta_a)\right\rangle
\Bigg) .
\end{equation}


\subsection*{Entropy production in the under-damped model}

Here we consider the under-damped model described in the main text:
\begin{align}
\begin{split}
d{r}_a^j =& \ {v}_a^j dt ,
\end{split}\\
\begin{split}
d{v}_a^j =& -\gamma{v}_a^j dt + \gamma v_0\hat{e}^j(\theta_a) dt - \frac{1}{m}\partial_{{r}^j_a}U_a(\mathbf{r}) dt + \sqrt{\frac{2\gamma}{\beta m}}dW_{{v}_a^j} ,
\end{split}\\
\begin{split}
d\theta_a =& \ \omega_a dt ,
\end{split}\\
\begin{split}
d\omega_a =& -\gamma_{\text{R}}\omega_a dt + \frac{1}{I}{\cal T}_a(\mathbf{r}, \theta) dt + \sqrt{\frac{2\gamma_{\text{R}}}{\beta I}}dW_{\omega_a} .
\end{split}
\end{align}
Regardless of the odd or even interpretation of $\theta$ under TRS, we have $A^{\text{IR}}_{\omega_a} {=} -\gamma_{\text{R}}\omega_a$, $A^{\text{REV}}_{\omega_a} {=} {\cal T}_a( \mathbf{r},\theta)/I$, $A^{\text{IR}}_{\theta_a} {=} 0$, $A^{\text{REV}}_{\theta_a} {=} \omega_a$, $A^{\text{IR}}_{{r}_a^j} {=} 0$, $A^{\text{REV}}_{{r}_a^j} {=}{v}_a^i$, $D_{\omega_a} {=} \gamma_\text{R}/\beta I$, $D_{\theta_a} {=} 0$, $D_{{v}_a^j} {=} \gamma/\beta m$, and $D_{{r}_a^j} {=} 0$.


\subsubsection*{Even self-propulsion}

For the even interpretation of $\theta$, we have $A^{\text{IR}}_{{v}_a^j} {=} - \gamma{v}_a^i$ and $A^{\text{REV}}_{{v}_a^j} {=} \gamma v_0\hat{e}^j(\theta_a) - \partial_{{r}^j_a}U_a(\mathbf{r})/m$.
Applying Eq.~\eqref{eq:ep} gives 
\begin{align}
d\Delta {S}^{\rm med}&=-\sum_j\beta\omega_j\circ d\omega_j+\beta\omega_j\mathcal{T}_j(\mathbf{r},\theta)dt\nonumber\\
&\qquad-\sum_{i=1}^2\beta m v_j^i\circ dv_j^i+\beta v^j_i(\partial_{r_j^i}U_j(\mathbf{r})-v_0\hat{e}^i(\theta_j))dt
\end{align}
Converting to It\={o} form, inserting the stochastic differentials $dv_j^i$ and $d\omega_j$ and taking expectations yields
\begin{equation}
\frac{d\Delta S^\text{med} }{dt}= \sum_{a=1}^N\Bigg(\gamma_\text{R}\Big(\beta I\left\langle\omega_a^2\right\rangle - 1\Big) + \sum_{j=1}^2 \gamma \bigg(\beta m \left\langle({v}^j_a)^2\right\rangle - 1 \bigg)\Bigg),
\label{eq:ep-noeq-odd}
\end{equation}
which may be straight forwardly decomposed into its per particle contributions. 
However, we may alternatively recognize that $\sum_{i,j}\langle v_j^i\circ dv_j^i\rangle=\langle dU(\mathbf{r})\rangle=0$ in the steady state, thus proceeding with the surviving terms to obtain
\begin{equation}
\frac{d\Delta S^\text{med} }{dt}= \sum_{a=1}^N\Bigg(\gamma_\text{R}\Big(\beta I\left\langle\omega_a^2\right\rangle - 1\Big) - \sum_{j=1}^2 m\gamma v_0\beta \langle \hat{e}^j(\theta_a)v_a^j\rangle\Bigg)
\end{equation}
allowing us to make the steady state connection
\begin{align}
\sum_{j,i}\langle \hat{e}^i(\theta_j)v_j^i\rangle &= \sum_{j,i}\frac{m\beta \langle (v_j^i)^2\rangle -1}{m\beta v_0}.
\label{connect}
\end{align}
\subsubsection*{Odd self-propulsion}

For the odd interpretation of $\theta$, we have $A^{\text{IR}}_{{v}_a^j} {=} - \gamma{v}_a^i + \gamma v_0\hat{e}^j(\theta_a)$ and $A^{\text{REV}}_{{v}_a^j} {=} -\partial_{{r}^j_a}U_a(\mathbf{r})/m$.
Applying Eq.~\eqref{eq:ep} we obtain:
\begin{align}
d\Delta {S}^{\rm med}&=-\sum_j\beta\omega_j\circ d\omega_j+\beta\omega_j\mathcal{T}_j(\mathbf{r},\theta)dt\nonumber\\
&\qquad+\sum_{i=1}^2\beta(-mv_j^i+mv_0\hat{e}^i(\theta_j))\circ dv_j^i+\beta(-v^j_i+v_0\hat{e}^i(\theta_j))\partial_{r_j^i}U_j(\mathbf{r})dt
\end{align}
Converting to It\={o} form, inserting the stochastic differentials $dv_j^i$ and $d\omega_j$ and taking expectations yields
\begin{equation}
d\Delta S^\text{med} = \sum_{a=1}^N\Bigg(\gamma_\text{R}\Big(\beta I\left\langle\omega_a^2\right\rangle - 1\Big) + \sum_{j=1}^2 \gamma\bigg(\beta m \left\langle(\mathbf{v}^j_a - v_0\hat{e}^j(\theta_a))^2\right\rangle - 1 \bigg)\Bigg),
\label{eq:ep-noeq-odd}
\end{equation}
which again forms a basis for a per particle contribution. 
Again, however, in the steady state we may assume $\sum_{i,j}\langle v_j^i\circ dv_j^i\rangle=\langle dU(\mathbf{r})\rangle=0$ such that we have
\begin{align}
d\langle\Delta {S}^{\rm med}\rangle&=\sum_j\beta\langle\omega \mathcal{T}_{j}(\mathbf{r},\theta)\rangle dt+m\beta\gamma v_0^2dt-\sum_{i=1}^2m\beta\gamma v_0\langle\hat{e}^i(\theta_j) v^i_j\rangle dt.
\end{align}
Utilizing Eq.~\eqref{connect} then gives
\begin{equation}
\frac{d\langle\Delta S^\text{med}\rangle}{dt} =
\sum_{a=1}^N
\Bigg(
\gamma_\text{R}\Big(\beta I\left\langle\omega_a^2\right\rangle - 1\Big)
+ \sum_{j=1}^2
\gamma \left(\frac{\beta mv_0^2}{2} - \beta m \left\langle({v}^j_a)^2\right\rangle + 1 \right)
\Bigg) 
\end{equation}
giving a starker contrast of the total contributions under odd and even interpretations of TRS for $\theta$.


\section*{Hidden entropy production beyond under-damped translational motion}
The entropy production formulae for the over-damped model immediately reduce to the free particle contributions reported in \cite{shankar2018hidden} upon setting $\partial_{r_a^i}U_a(\mathbf{r})=\mathcal{T}_{a}(\mathbf{r},\theta)=0$. For a single particle, the underdamped model under the same conditions for odd $\theta$ we have
\begin{align}
d\langle\Delta {S}^{\rm med}\rangle&=m\beta\gamma v_0^2dt-\sum_{i=1}^2m\beta\gamma v_0\langle\hat{e}^i(\theta) v^i_j\rangle dt.
\end{align}
 and for even $\theta$
\begin{align}
d\langle\Delta {S}^{\rm med}\rangle&=\sum_{i=1}^2m\beta\gamma v_0\langle\hat{e}^i(\theta) v^i\rangle dt.
\end{align}

Thus explicit expressions depend on determining the steady state correlation $\langle\hat{e}^i(\theta) v^i\rangle$. We can calculate this explicitly in the zero inertia limit for the dynamics of $\theta$. Under such conditions we may write
\begin{align}
d\theta&=\sqrt{2D_{\text{R}}} dW_{\theta}
\end{align}
where $D_{\text{R}}= (\gamma_{\text{R}} I\beta)^{-1}$. 
By It\={o}'s lemma we have
\begin{align}
d\hat{e}^1=-D_{\text{R}}\hat{e}^1dt-\sqrt{2D_{\text{R}}(1-(\hat{e}^1)^2)} dW_{\theta}
\end{align}
with analogous expression for $d\hat{e}^2$. This has integrating factor solution
\begin{align}
\hat{e}^1(t)=\hat{e}^1(0)e^{-D_{\text{R}} t}-\int_{0}^t e^{-D_{\text{R}}(t-t')}\sqrt{1-(\hat{e}^1(t'))^2}\sqrt{2D_{\text{R}}} dW_{\theta}(t').
\end{align}
Similarly, we have an integrating factor solution for $v_x(t)$
\begin{align}
v^1(t)=v^1(0)e^{-\gamma t}+\int_{0}^te^{-\gamma(t-t')}\gamma v_0 \hat{e}^1(t')dt'+\int_{0}^te^{-\gamma(t-t')}\sqrt{\frac{2\gamma}{m\beta}}dW_{v^1}(t')
\label{vsol}
\end{align}
so
\begin{align}
\langle \hat{e}^1(t_e)v^1(t_v)\rangle &=\Bigg\langle\left(\hat{e}^1(0)e^{-D_{\text{R}} t_e}-\int_{0}^{t_e} e^{-D_{\text{R}}(t_e-t_e')}\sqrt{2D_{\text{R}}(1-(\hat{e}^1(t_e'))^2)} dW_{\theta}(t_e')\right)\nonumber\\
&\quad\times\left(v^1(0)e^{-\gamma t_v}+\int_{0}^{t_v}e^{-\gamma(t_v-t_v')}\gamma v_0 \hat{e}^1(t_v')dt_v'+\int_{0}^{t_v}e^{-\gamma(t_v-t_v')}\sqrt{\frac{2\gamma}{m\beta}}dW_{v^1}(t_v')\right)\Bigg\rangle.
\end{align}
or  explicitly writing $\hat{e}^1(t_v')$
\begin{align}
&\langle \hat{e}^1(t)v^1(t)\rangle =\Bigg\langle\left(\hat{e}^1(0)e^{-D_{\text{R}} t}-\int_{0}^t e^{-D_{\text{R}}(t-t')}\sqrt{1-(\hat{e}^1(t'))^2}\sqrt{2D_{\text{R}}} dW_{\theta}(t')\right)\nonumber\\
&\times\Bigg(v^1(0)e^{-\gamma t_v}+\int_{0}^{t_v}e^{-\gamma(t_v-t_v')}\gamma v_0 \left[\hat{e}^1(0)e^{-D_{\text{R}}t_v'}-\int_0^{t_v'}e^{-D_{\text{R}}(t_v'-t_v'')}\sqrt{2D_{\text{R}}}\sqrt{1-(\hat{e}^1(t_v''))^2}dW_{\theta}(t'')\right]dt_v'\nonumber\\
&\qquad+\int_{0}^{t_v}e^{-\gamma(t_v-t_v')}\sqrt{\frac{2\gamma}{m\beta}}dW_{v^1}(t_v')\Bigg)\Bigg\rangle.
\end{align}
We compute this in the $t\to\infty$ limit corresponding to the steady state. When we do this all terms will disappear, either through the averaging, i.e.$\langle dW_i\rangle =0$ or through vanishing exponentials, except the term that contains $dW_{\theta}dW_{\theta}$ which we write as
\begin{align}
\left\langle\gamma v_0\int_0^{t_e}\int_0^{t_v}\int_0^{t'_v}e^{-\gamma(t_v-t'_v)}e^{-D_{\text{R}}(t_e-t'_e)}e^{-D_{\text{R}}(t'_v-t''_v)}2D_{\text{R}}\sqrt{1-(\hat{e}^1(t'_e))^2}\sqrt{1-(\hat{e}^1(t''_v))^2}dW_{\theta}(t'_e)dW_{\theta}(t''_v)dt'_v\right\rangle
\end{align}
Sifting out with the delta correlated Wiener processes, i.e.
\begin{align}
\int_0^t\int_0^{t'} f(t'',t''')dW_{\theta}(t'')dW_{\theta}(t''')&=\int_0^t\int_0^{t'} f(t'',t''')\delta(t''-t''')dt'' dt'''\nonumber\\
&=\int_0^{t'} f(t'',t'')dt'' \qquad(t>t')
\end{align}
and considering $t_e=t_v$ such that $t_e>t'_v$ we write
\begin{align}
\left\langle\gamma v_0\int_0^{t_v}\int_0^{t'_v}e^{-\gamma(t_v-t'_v)}e^{-D_{\text{R}}(t_e-t''_v)}e^{-D_{\text{R}}(t'_v-t''_v)}2D_{\text{R}}(1-(\hat{e}^1(t''_v))^2)dt''_v dt'_v\right\rangle
\end{align}
which becomes
\begin{align}
\gamma v_0\int_0^{t_v}\int_0^{t'_v}e^{-\gamma(t_v-t'_v)}e^{-D_{\text{R}}(t_e-t''_v)}e^{-D_{\text{R}}(t'_v-t''_v)}2D_{\text{R}}(1-\langle(\hat{e}^1(t''_v))^2\rangle)dt''_v dt'_v.
\end{align}
In the $t_e=t_v\to\infty$ limit we may safely write $\langle(\hat{e}^1(t''_v))^2\rangle=1/2$ and thus
\begin{align}
\gamma v_0\int_0^{t_v}\int_0^{t'_v}e^{-\gamma(t_v-t'_v)}e^{-D_{\text{R}}(t_e-t''_v)}e^{-D_{\text{R}}(t'_v-t''_v)}D_{\text{R}}dt''_v dt'_v.
\end{align}
Computing the integral and setting $t_e=t_v=t$ we find
\begin{align}
\langle \hat{e}^1(t)v^1(t)\rangle&=\gamma v_0e^{-D_{\text{R}}t}\frac{D_{\text{R}}\cosh(D_{\text{R}} t)-\gamma\sinh(D_{\text{R}} t)-D_{\text{R}}e^{-\gamma t}}{(D_{\text{R}}+\gamma)(D_{\text{R}}-\gamma)}.
\end{align}
Considering the $t\to\infty$ limit then gives
\begin{align}
\langle \hat{e}^1(t)v^1(t)\rangle&=\frac{\gamma v_0}{2(D_{\text{R}}+\gamma)}
\label{xp}
\end{align}
and so by symmetry
\begin{align}
\sum_{i=1}^2\langle \hat{e}^i(t)v^i(t)\rangle&=\frac{\gamma v_0}{(D_{\text{R}}+\gamma)}
\end{align} 
and thus for odd $\theta$ gives
\begin{align}
d\langle\Delta \mathcal{S}_{\rm med}\rangle&=m\beta\gamma v_0^2dt-\sum_{i=1}^2 \langle v^i(t)\hat{e}^i(t)\rangle m \beta\gamma v_0dt\nonumber\\
&=m\beta \gamma v_0^2dt-\frac{m\beta \gamma^2 v^2_0}{D_{\text{R}}+\gamma}dt\nonumber\\
&=\frac{m\beta\gamma D_{\text{R}} v_0^2}{D_{\text{R}}+\gamma}dt\nonumber\\
&=\frac{m\gamma \beta v_0^2}{1+\gamma\gamma_{\text{R}}\beta I}dt
\end{align}
and for even $\theta$ gives
\begin{align}
d\langle\Delta \mathcal{S}_{\rm med}\rangle&=\sum_{i=1}^2 \langle v^i(t)\hat{e}^i(t)\rangle m \beta\gamma v_0dt\nonumber\\
&=\frac{m\beta \gamma^2 v^2_0}{D_{\text{R}}+\gamma}dt\nonumber\\
&=\frac{m\gamma^2 \beta^2 I\gamma_{\text{R}} v_0^2}{1+\gamma\gamma_{\text{R}}\beta I}dt
\end{align}
in agreement with \cite{shankar2018hidden}.
\par
The above, however, relies upon an over-damped description for the dynamics in $\theta$, not consistent with the full under-damped equation of motion. Direct computation of $\langle e^1(t)v^1(t)\rangle$ as above suffers from the non-linearity of the transform of the unit vector. However, we can utilize Eq.~(\ref{connect}) to consider instead the long term variance $\langle (v^1(t)^2)\rangle$, so as to exploit Eq.~(\ref{connect}), which we can construct using the same integrating factor solution in Eq.~(\ref{vsol}). Expanding, taking the limit $t\to\infty$, neglecting terms that vanish in the limit and taking expectations such that first order integrals in Wiener processes vanish leaves
\begin{align}
\langle (v^1(t)^2)\rangle&=\lim_{t\to\infty}\frac{2\gamma}{m\beta}\int_0^t \int_0^t e^{-\gamma(2t-t_1-t_2)} \langle dW_{v^1}(t_1)dW_{v^1}(t_2) \rangle \nonumber\\
&\quad+\lim_{t\to\infty}\gamma^2v_0^2\int_0^t dt_1\int_0^t dt_2e^{-\gamma(2t-t_1-t_2)}\langle e^1(t_1)e^2(t_2)\rangle\nonumber\\
&=\frac{1}{m\beta}+\lim_{t\to\infty}\gamma^2v_0^2\int_0^t dt_1\int_0^t dt_2e^{-\gamma(2t-t_1-t_2)}\langle e^1(t_1)e^2(t_2)\rangle\\
&=\frac{1}{m\beta}+\frac{\gamma^2v_0^2\mathcal{G}}{2},
\end{align}
defining $\mathcal{G}$. When paired with Eq.~(\ref{connect}) this gives, for odd $\theta$,
\begin{align}
\langle\Delta\dot{S}^{\text{tot}}\rangle&=m \beta\gamma v_0^2(1-\gamma^2\mathcal{G})
\end{align}
and for even $\theta$
\begin{align}
\langle\Delta\dot{S}^{\text{tot}}\rangle&=m \beta\gamma^3 v_0^2\mathcal{G}.
\end{align}
When $\theta$ is described by over-damped equations of motion then $\langle e^1(t_1)e^2(t_2)\rangle_{\text{over}}=(1/2)e^{-(I\beta\gamma_{\text{R}})^{{-}1}|t_1{-}t_2|}$ such that $\mathcal{G}=\mathcal{G}_{\text{over}}=(\gamma(\gamma+(\beta I\gamma_{\text{R}})^{-1}))^{-1}$ also in agreement with the above.
\par
However, $\langle e^1(t_1)e^2(t_2)\rangle$ differs under an under-damped description. To find such a form we first consider $\langle (\theta(t_2)-\theta(t_1))^2\rangle$. First we integrate to find $\theta(t)$
\begin{align}
\theta(t)&=\theta(0)+\omega(0)\int_0^tdt'\;e^{-\gamma_{\text{R}}t'}+\sqrt{\frac{2\gamma_{\text{R}}}{I\beta}}\int_0^tdt'\int_0^{t'}dW(t'')\;e^{-\gamma_{\text{R}}(t'-t'')}\nonumber\\
&=\theta(0)+\frac{\omega(0)}{\gamma_{\text{R}}}(1-e^{-\gamma_{\text{R}}t})+\sqrt{\frac{2\gamma_{\text{R}}}{I\beta}}\int_0^{t}dW(t'')\int_{t''}^tdt'\;e^{-\gamma_{\text{R}}(t'-t'')}\nonumber\\
&=\theta(0)+\frac{\omega(0)}{\gamma_{\text{R}}}(1-e^{-\gamma_{\text{R}}t})+\sqrt{\frac{2}{I\beta \gamma_{\text{R}}}}\int_0^{t}(1-e^{-\gamma_{\text{R}}(t-t'')})dW(t''),
\end{align}
from which we obtain
\begin{align}
\langle (\theta(t)-\theta(0))^2\rangle &=\frac{\langle\omega^2(0)\rangle}{\gamma^2_{\text{R}}}(1-e^{-\gamma_{\text{R}}t})^2+\frac{2}{I\beta \gamma_{\text{R}}}\int_0^{t}(1-e^{-\gamma_{\text{R}}(t-t')})^2 dt'\nonumber\\
&=\frac{\langle\omega^2(0)\rangle}{\gamma^2_{\text{R}}}(1-e^{-\gamma_{\text{R}}t})^2+\frac{2}{I\beta \gamma_{\text{R}}}\frac{2\gamma_{\text{R}}t-3-e^{-2\gamma_{\text{R}}t}+4e^{-\gamma_{\text{R}}t}}{2\gamma_{\text{R}}}.
\end{align}
With $\langle \omega^2(0)\rangle = (\beta I)^{-1}$ corresponding to the steady state we have
\begin{align}
\langle (\theta(t)-\theta(0))^2\rangle &=\frac{2}{\gamma_{\text{R}}^2\beta I}(\gamma_{\text{R}}t-1+e^{-\gamma_{\text{R}}t}),\;t>0.
\end{align}
Expecting time translation invariance and symmetry we then have
\begin{align}
\langle (\theta(t_2)-\theta(t_1))^2\rangle &=\frac{2}{\gamma_{\text{R}}^2\beta I}(\gamma_{\text{R}}|t_2-t_1|-1+e^{-\gamma_{\text{R}}|t_2-t_1|}).
\end{align}
Crucially, a theorem of centered Gaussian variables states \cite{william_t_coffey_langevin_2012}
\begin{align}
\langle \cos(\theta(t_2))\cos(\theta(t_1))\rangle &=\frac{1}{2}e^{-\frac{1}{2}\langle (\theta(t_2)-\theta(t_1))^2\rangle}
\end{align}
such that we finally have (introducing subscripts for the nature of the \emph{rotational} dynamics)
\begin{align}
\langle e^1(t_1)e^2(t_2)\rangle_{\text{under}}&=\frac{1}{2}e^{-({\gamma_{\text{R}}^2\beta I})^{-1}(\gamma_{\text{R}}|t_2-t_1|-1+\exp[{-\gamma_{\text{R}}|t_2-t_1|}])}\nonumber\\
&=\langle e^1(t_1)e^2(t_2)\rangle_{\text{over}}\exp[({\gamma_{\text{R}}^2\beta I})^{-1}(1-e^{-\gamma_{\text{R}}|t_2-t_1|}])
\end{align}
revealing that $\mathcal{G}_{\text{under}}\geq\mathcal{G}_{\text{over}}$ and thus for odd and even $\theta$, $\langle\Delta\dot{S}^{\rm tot}\rangle_{\text{under}}\leq \langle\Delta\dot{S}^{\rm tot}\rangle_{\text{over}}$ and $\langle\Delta\dot{S}^{\rm tot}\rangle_{\text{under}}\geq \langle\Delta\dot{S}^{\rm tot}\rangle_{\text{over}}$ respectively for stationary free ABP dynamics with under-damped translational motion.
\par
That in the steady state for a given TRS interpretation for $\theta$, coarse-graining in the rotational dynamics causes an over or under-estimation of the entropy production for any system parameters leads to the claim of a hidden entropy production associated with such coarse-graining.

The integral for $\mathcal{G}_{\text{under}}$ generally has no close form solution, but we can calculate it in the case of all free parameters set to $1$.
In this case, it reads
\begin{align}
\mathcal{G}_{\text{under}} &= \lim_{t\to\infty}\int_{0}^{t}dt_1\int_{0}^{t}dt_2 e^{-(2t-t_1-t_2)}e^{-|t_1-t_2|+1-\exp{(-|t_1-t_2|)}}\nonumber\\
&= \lim_{t\to\infty}e^{-2t}\left(-2e^{2t}+e^{1+t-\cosh(t)+\sinh(t)}(1+e^t)-e\text{Ei}(-1)+e\text{Ei}(-\cosh(t)+\sinh(t))\right)\nonumber\\
&=e-2.
\end{align}
For the same parameters $\mathcal{G}_{\text{over}}=1/2$ and so the ratio $\mathcal{G}_{\text{under}}/\mathcal{G}_{\text{over}}=2e-4\simeq 1.44$, leading to the the ratios $\langle\Delta \dot{S}^\text{tot}\rangle_{\text{under}}/\langle\Delta \dot{S}^\text{tot}\rangle_{\text{over}}=2e-4\simeq 1.44$ for even $\theta$ and $\langle\Delta \dot{S}^\text{tot}\rangle_{\text{under}}/\langle\Delta \dot{S}^\text{tot}\rangle_{\text{over}}=6-2e\simeq 0.563$ for odd $\theta$.
\section*{Numerical Integration}
We utilize a stochastic velocity Verlet algorithm described in detail in \cite{kloeden_numerical_1992}. For each particle, indexed by $a$, there are three momenta variables $\omega_a$, $v_a^1$ and $v_2^2$. Each such variable requires two zero mean, unit variance, Guassian distributed pseudo-random numbers, written $\phi_a^{1,1}, \phi_a^{1,2}, \phi_a^{2,1}, \phi_a^{2,2}, \phi_a^{\theta,1}, \phi_a^{\theta,2}$, which are all mutually independent (i.e. $\langle \phi_a^{x,y}\phi^{m,n}_b\rangle =\delta_{ab}\delta_{x,m}\delta_{y,n}$). Recalling $\mathbf{r}=\{\mathbf{r}_1,\ldots,\mathbf{r}_N\}$ (with similarly defined $\mathbf{v},\mathbf{\theta},\mathbf{\omega}$) and $\mathbf{r}_a=\{{r}_a^1,{r}_a^2\}$, $\mathbf{v}_a=\{{v}_a^1,{v}_a^2\}$, the algorithm then reads
\begin{align}
C_a^1(t)&=\frac{(\Delta t)^2}{2}\left[m^{-1}\mathcal{F}_a^1(\mathbf{r}(t),\theta(t))-\gamma v_a^1(t)\right]+\sqrt{\frac{\gamma}{m\beta}}\frac{(\Delta t)^{3/2}}{2}\left(\phi_a^{1,1}+\frac{\phi_a^{1,2}}{\sqrt{3}}\right)\nonumber\\
C_a^2(t)&=\frac{(\Delta t)^2}{2}\left[m^{-1}\mathcal{F}_a^2(\mathbf{r}(t),\theta(t))-\gamma v_a^2(t)\right]+\sqrt{\frac{\gamma}{m\beta}}\frac{(\Delta t)^{3/2}}{2}\left(\phi_a^{2,1}+\frac{\phi_a^{2,2}}{\sqrt{3}}\right)\nonumber\\
C_a^\theta(t)&=\frac{(\Delta t)^2}{2}\left[I^{-1}\mathcal{T}_a(\mathbf{r}(t),\theta(t))-\gamma_{\text{R}} \omega_a(t)\right]+\sqrt{\frac{\gamma_{\text{R}}}{I\beta}}\frac{(\Delta t)^{3/2}}{2}\left(\phi_a^{\theta,1}+\frac{\phi_a^{\theta,2}}{\sqrt{3}}\right)\nonumber\\
r_a^1(t+\Delta t)&=r_a^1(t)+v_a^1(t)\Delta t +C_a^1(t)\nonumber\\
r_a^2(t+\Delta t)&=r_a^2(t)+v_a^2(t)\Delta t +C_a^2(t)\nonumber\\
\theta_a(t+\Delta t)&=\theta_a(t)+\omega_a(t)\Delta t +C_a^\theta(t)\nonumber\\
v_a^1(t+\Delta t)&=v_a^1(t)-\gamma v_a^1(t)\Delta t +\frac{\Delta t}{2m}\left[\mathcal{F}_a^1(\mathbf{r}(t),\theta(t))+\mathcal{F}_a^1(\mathbf{r}(t+\Delta t),\theta(t+\Delta t))\right]-\gamma C_a^1(t)+\sqrt{\frac{\gamma \Delta t}{m\beta}}\phi_a^{1,1}\nonumber\\
v_a^2(t+\Delta t)&=v_a^2(t)-\gamma v_a^2(t)\Delta t +\frac{\Delta t}{2m}\left[\mathcal{F}_a^2(\mathbf{r}(t),\theta(t))+\mathcal{F}_a^2(\mathbf{r}(t+\Delta t),\theta(t+\Delta t))\right]-\gamma C_a^2(t)+\sqrt{\frac{\gamma \Delta t}{m\beta}}\phi_a^{2,1}\nonumber\\
\omega_a(t+\Delta t)&=\omega_a(t)-\gamma \omega_a(t)\Delta t +\frac{\Delta t}{2I}\left[\mathcal{F}_a^\theta(\mathbf{r}(t),\theta(t))+\mathcal{F}_a^\theta(\mathbf{r}(t+\Delta t),\theta(t+\Delta t))\right]-\gamma C_a^\theta(t)+\sqrt{\frac{\gamma_{\text{R}} \Delta t}{I\beta}}\phi_a^{\theta,1}.\nonumber
\end{align}
These equations were simulated using $\Delta t =0.008$.
\bibliographystyle{apsrev4-1}
%

\end{widetext}

\end{document}